\documentstyle[12pt,twoside,fleqn,espcrc1,epsf]{article}
\title{$S$-matrix studies of resonances in $A$ = 3, 4, 5, 6, and 
12 nucleon systems}
\author{Attila Cs\'ot\'o$^{\rm a}$ and G.~M. Hale\address{Theoretical 
Division, Los Alamos National Laboratory, Los Alamos, NM 87545, USA}}

\begin{document}
\maketitle 

\begin{abstract}
Resonances of certain light nuclei are explored by studying the
complex pole structures of the scattering matrices. Among other results 
we predict the existence of three-neutron and three-proton resonances, 
a small spin-orbit splitting in the low-lying $^5$He and $^5$Li states 
and the nonexistence of the soft dipole resonance in $^6$He.
\end{abstract}

\section{Introduction}
The investigation of resonance structures in light nuclei offers a 
rich source of information on many-body dynamics, nucleon-nucleon
interaction, shell-structure, etc. Conventionally, resonances are
studied both experimentally and theoretically by analyzing certain
observables (cross sections, scattering amplitudes, phase shifts,
etc.) at {\it real energies} using the methods and concepts of
scattering theory. However, in scattering theory resonances are
defined through the analytic properties of certain quantities ($S$
matrix, Fredholm determinant, Jost function) at {\it complex
energies}. We believe that the theoretical study of few-nucleon
scattering at complex energies offers a unique insight into the
dynamics of these problems. 

We study selected resonances of $A$ = 3, 4, 5, 6, and 12 nucleon 
systems by exploring the analytic properties of the $S$ matrices 
at complex energies. The nuclei are described within the Resonating 
Group Model (RGM), assuming 2- or 3-cluster dynamics. Certain 
$S$-matrix poles are also extracted from $R$-matrix fits of 
experimental data by extending the $S$ matrix, generated from the 
$R$ matrix, to complex energies. 

\section{Localization of S-matrix poles at complex energies}
We use a microscopic two/three-cluster description of the
$^3n(=n+n+n)$, $^3p(=p+p+p)$, $^4{\rm He}(=\{t+p,h+n\})$, $^5{\rm
He}(=\alpha+n)$, $^5{\rm Li}(=\alpha+p)$, $^6{\rm
He}(=\alpha+n+n)$, $^6{\rm Li}(=\alpha+p+n)$, $^6{\rm 
Be}(=\alpha+p+p)$, and $^{12}{\rm C}(=\alpha+\alpha+\alpha)$
nuclei. Here $\alpha={^4{\rm He}}$, $t={^3{\rm H}}$, $h={^3{\rm
He}}$, and the cluster structures, assumed in the model, are
indicated. We assume simple $0s$ harmonic oscillator shell-model 
wave functions for the internal states of the clusters ($\alpha$, 
$t$, and $h$). However, the relative motions between the clusters, 
which are the most important degrees of freedom, are treated with 
the rigor of few-body physics. 

The wave function of a two- and three-cluster system looks like
\begin{equation}
\Psi=\sum_{L,S}{\cal A} \Bigg \{ \bigg [ \Big [\Phi^{A}
\Phi^{B} \Big ]_S\chi_L(\mbox{\boldmath $\rho$})
\bigg ]_{JM} \Bigg\}, 
\label{wfn2}
\end{equation}
and
\begin{equation}
\Psi=\sum_{l_1,l_2,L,S}{\cal A} \Bigg \{ \bigg [ 
\Big [\Phi^{A} \Phi^{B} \Phi^{C}\Big ]_S
\chi_{[l_1,l_2]L}(\mbox{\boldmath $\rho$}_1,
\mbox{\boldmath $\rho$}_2)\bigg ]_{JM} \Bigg\}, 
\label{wfn3}
\end{equation}
respectively. Here ${\cal A}$ is the intercluster antisymmetrizer, 
the $\Phi$ cluster internal states are translationally invariant 
$0s$ harmonic-oscillator shell-model states, the $\mbox{\boldmath 
$\rho$}$ vectors are the intercluster relative coordinates, 
$l_1$ and $l_2$ are the angular momenta of the two relative 
motions, $L$ is the total orbital angular momentum, $S$ is the
total spin, and $[\ldots]$ denotes angular momentum coupling. In 
the case of three-cluster dynamics all possible sets of relative 
coordinates [$A(BC)$, $C(AB)$, $B(AC)$] and angular momentum 
couplings are included in (\ref{wfn3}).

Putting (\ref{wfn2}) or (\ref{wfn3}) into the $N$-body
Schr\"odinger equation we get equations for the unknown relative
motion functions $\chi$. For two-body (three-body) bound states
they are expanded in terms of (products of) Gaussian functions, 
and the expansion coefficients are determined from a variational
principle for the energy. For two-body scattering states the $\chi$
functions are expanded in terms of Gaussian functions matched with
the correct asymptotics, and the expansion coefficients are
determined from the Kohn-Hulth\'en variational method for the $S$
matrix \cite{Kamimura}.

In scattering theory resonances are defined as complex-energy
solutions of the Schr\"odin\-ger equation that correspond to the
poles of the $S$ matrix (or equivalently the zeros of the Fredholm
determinant or Jost function). In order to obtain these complex
solutions, we implemented a direct analytic continuation of the $S$
matrix for two-cluster systems \cite{A4,A5} and the complex scaling
method for three-cluster systems \cite{A3,A6,A12}. 

For two-cluster systems we solve the Schr\"odinger equation for 
the relative motion at complex energies with the the following
boundary condition for $\rho\rightarrow \infty$ 
\begin{equation}
\chi(\varepsilon,\rho) 
\rightarrow H^-(k\rho)-\tilde S(\varepsilon) H^+(k\rho).
\end{equation}
Here $\varepsilon$ and $k$ are the {\it complex} energies
and wave numbers of the relative motions, and $H^-$ and 
$H^+$ are the incoming and outgoing Coulomb functions, 
respectively. The function $\tilde S$ has no physical 
meaning, except if it is singular at the energy 
$\varepsilon$. Then $\tilde S$ coincides with the physical 
$S$ matrix describing a purely outgoing solution, that is a 
resonance. So we search for the poles of $\tilde S$ at complex 
energies and extract the resonance parameters from 
$\varepsilon=E_{\rm r}-i\Gamma/2$ \cite{pole}.

For the three-cluster systems we solve the eigenvalue problem of a
new Hamiltonian defined by
\begin{equation}
\widehat{H}_\theta=\widehat{U}(\theta)\widehat{H}
\widehat{U}^{-1}(\theta),
\end{equation}
where $\widehat{H}$ is the original many-body Hamiltonian and
$\widehat{U}$ is the complex scaling transformation which acts on a
function $f({\bf r})$ as $\widehat{U}(\theta)f({\bf r})=e^{3 i 
\theta /2}f({\bf r}e^{i\theta})$.
In the case of a multicluster system the transformation
is performed on each dynamical coordinate (relative motion). The
solution of the complex-scaled Schr\"odinger equation 
results in a spectrum with continuum cuts rotated by
$2\theta$ relative to the real energy axis plus possibly a few 
isolated complex points at the resonance and bound state poles 
\cite{CS}.

We can perform a similar analysis of the analytic properties of the
$S$ matrices in a model-independent way starting from the
experimental data. The data can be described by the usual
$R$-matrix method that works at real energies. Then, an $S$ matrix
can be constructed from this $R$ matrix and can be analytically
continued to complex energies, and its poles can be localized 
\cite{Rmat}. We call this procedure the extended $R$-matrix method. 
So far we have implemented this method for two-cluster dynamics.

\section{Application to A = 3, 4, 5, 6, and 12 nucleon systems}
We briefly summarize the physics motivation and the main results of
Refs.\ \cite{A4,A5,A3,A6,A12}, where the methods, discussed above,
were applied to selected resonances of various light nuclei.
Further details can be found in the original papers. In all of our
RGM calculations we used the Minnesota effective $NN$ interaction 
\cite{MN} which provides a reasonably good overall description of 
the low-energy $N+N$ scattering phase shifts \cite{A3} and the bulk
properties of the $^3$H, $^3$He, and $^4$He clusters.

\begin{figure}
\begin{minipage}[t]{10cm}
\epsfxsize=9cm \epsfbox{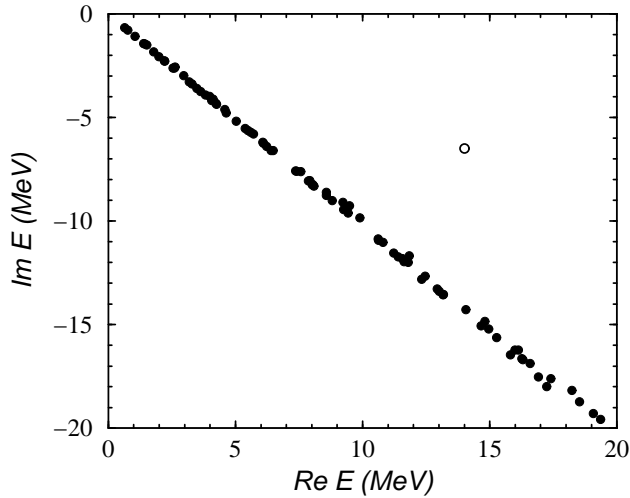}
\end{minipage}
\hspace{\fill}
\begin{minipage}[b]{6cm}
\caption{Energy eigenvalues of the complex scaled Hamiltonian 
for $3/2^+$ three-neutron states. The dots are the points of the 
rotated discretized continuum, while the circle is a three-neutron 
resonance. $E_r={\rm Re}(E)$, $\Gamma=-2\,{\rm Im}(E)$.}
\end{minipage}
\end{figure}

{\it A=3}: We searched for three-neutron resonances \cite{A3}, 
which had been predicted from pion double charge exchange 
experiments on $^3$He. All partial waves up to $J=5/2$ were 
nonresonant except the $J^\pi=3/2^+$ one, where we found a state 
at $E=14$ MeV energy with 13 MeV width. The parameters of the 
mirror state in the three-proton system are $E=15$ MeV and 
$\Gamma=14$ MeV. Fig.\ 1 shows the energy eigenvalues of the 
complex scaled Hamiltonian for $J^\pi=3/2^+$ three-neutron states. 
The three-body continuum is rotated by $2\theta$ and is discretized 
because of our finite variational basis. The position of the 
three-neutron resonance, isolated from the continuum line, is 
approximately independent of $\theta$. We have also begun a 
charge-independent $R$-matrix analysis of $N+d$ data at energies 
below the breakup threshold \cite{GMH}. Preliminary indications 
from the analysis are that a number of $S$-matrix poles exist is 
the $S$-wave and $P$-wave states, some of which are (subthreshold) 
virtual resonances.

{\it A=4}: We studied the $0^+_2$ state of $^4$He \cite{A4}, which 
state is notoriously difficult to reproduce in shell-models. In a 
$\{^3{\rm H}+p,$$^3{\rm He}+n\}$ model, which reproduced well the 
relevant $^1S_0$ $^3{\rm H}+p$ phase shift, an $S$-matrix pole was 
found corresponding to $E_r=93$ keV and $\Gamma=390$ keV resonance 
parameters. This state was also localized with $E_r=114$ keV and 
$\Gamma=392$ keV parameters in an $S$ matrix that was constructed 
from a comprehensive $R$-matrix fit to the data. Both calculations 
agree that this is the first excited state of $^4$He, and is a 
conventional resonance between the $^3{\rm H}+p$ and 
$^3{\rm He}+n$ thresholds.

\begin{figure}
\begin{minipage}[t]{10cm}
\epsfxsize=9cm \epsfbox{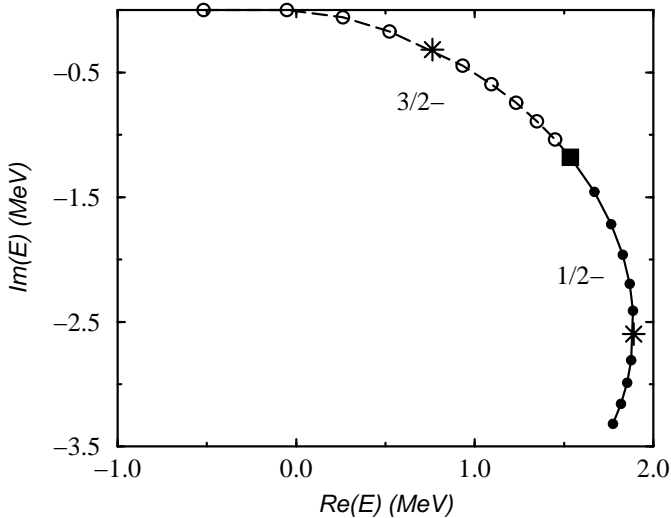}
\end{minipage}
\hspace{\fill}
\begin{minipage}[b]{6cm}
\caption{Trajectories of the $3/2^-$ and $1/2^-$ poles of $^5$He 
as a function of the spin-orbit strength. See the text for the
explanation of the symbols.}
\end{minipage}
\end{figure}

{\it A=5}: We studied the low-energy $^5$He and $^5$Li resonances 
\cite{A5} with the aim to resolve the apparent contradiction 
between the experimentally recommended resonance parameters and 
the ones that appear, e.g., in the neutron halo studies of $^6$He. 
Our complex $S$-matrix resonance parameters coming from an 
$\alpha+N$ RGM model and from an extended $R$-matrix model, 
respectively, were found to be in good agreement with each other. 
However, they differed from the conventional parameters, extracted 
from reaction cross sections at real energies. The most striking 
disagreement with the currently accepted resonance parameters 
\cite{Ajzenberg} is in the spin-orbit splitting of the $3/2^-$ and 
$1/2^-$ states. Our calculations predict a much smaller splitting 
than given in \cite{Ajzenberg}. In order to demonstrate the 
dependence of the $3/2^-$ and $1/2^-$ $S$-matrix pole positions on 
the spin-orbit strength ($V_{\rm SO}$) we show the $^5$He pole 
trajectories in Fig.\ 2. The solid square shows the result for 
$V_{\rm SO}=0$, while the stars correspond to the physical value 
of the strength. By further increasing $V_{\rm SO}$, the $3/2^-$ 
state becomes bound. One can see that the behavior of the pole 
trajectory rules out the possibility that the spin-orbit splitting 
is larger than 2 MeV.

{\it A=6}: The low-lying three-body resonances of $^6$He, $^6$Li, 
and $^6$Be were studied in an $\alpha+N+N$ RGM model \cite{A6}. 
Our motivation was to confirm or refute the existence of a new type 
of collective excitations, the soft dipole resonance, in the 
neutron halo nucleus $^6$He. This soft dipole resonance was
supposed to arise from the oscillation of the halo neutrons against
the $^4$He core. As the coupling between the core and the halo is
weak, this would result in a low-energy state compared to the usual
giant dipole resonances. We found all experimentally known 
three-body resonances of the $A=6$ nuclei, but no indication for 
the existence of a $1^-$ state in $^6$He. Thus, using one of the 
most comprehensive models of that nucleus, we showed that the soft 
dipole resonance does not exist in $^6$He. This result has been 
confirmed by other calculations in different models \cite{soft} 
and by a recent experiment \cite{soft1}.

{\it A=12}: We studied the low-lying natural-parity three-alpha 
resonances in $^{12}$C in order to shed light on the nature of the 
$0^+_2$ state \cite{A12}. This state, which plays an important role 
in stellar nucleosynthesis, was believed to be a three-alpha 
resonance. However, there was no unambiguous proof to support this 
claim. We localized all known low-energy natural-parity states of 
$^{12}$C by using $NN$ forces which correctly reproduced the 
$^8$Be ground state resonance. We unambiguously showed for the 
first time in a microscopic model, that the $0^+_2$ state is really 
a three-alpha resonance.

\section{Conclusion}

In summary, our study of resonances in light nuclei by means of the
exploration of the analytic structure of the $S$ matrix at complex
energies has proven to be very fruitful. We observe a good general
agreement between the results of our RGM and extended $R$-matrix
models. In contrast, resonance prescriptions formulated
at real energies, can lead to quite different results from one
another, especially for broad states. Thus, we recommend using the 
complex $S$-pole prescription to specify resonance parameters 
in all cases.

\mbox{}

This work was performed under the auspices of the U.S. Department
of Energy.

\end{document}